\documentclass[onecolumn, 9pt]{article}
\usepackage{extsizes}
\usepackage[super,sort&compress,comma]{natbib} 
\usepackage[version=3]{mhchem}
\usepackage[left=1.5cm, right=1.5cm, top=1.785cm, bottom=2.0cm]{geometry}
\usepackage{balance}
\usepackage{mathptmx}
\usepackage{sectsty}
\usepackage{comment}
\usepackage{graphicx} 
\usepackage{lastpage}
\usepackage[format=plain,justification=justified,singlelinecheck=false,font={stretch=1.125,small,sf},labelfont=bf,labelsep=space]{caption}
\usepackage{float}
\usepackage{fancyhdr}
\usepackage{fnpos}
\usepackage[english]{babel}
\addto{\captionsenglish}{%
  
}
\def\la{\langle}
\def\ra{\rangle}
\usepackage{array}
\usepackage{droidsans}
\usepackage{charter}
\usepackage[T1]{fontenc}
\usepackage[usenames,dvipsnames]{xcolor}
\usepackage{setspace}
\usepackage[compact]{titlesec}
\usepackage{hyperref}
\usepackage{lipsum}

\usepackage{epstopdf}

\definecolor{cream}{RGB}{222,217,201}

\begin{document}

\thispagestyle{plain}

\makeFNbottom
\makeatletter
\renewcommand\LARGE{\@setfontsize\LARGE{12pt}{17}}
\renewcommand\Large{\@setfontsize\Large{12pt}{14}}
\renewcommand\large{\@setfontsize\large{10pt}{12}}
\renewcommand\footnotesize{\@setfontsize\footnotesize{7pt}{10}}
\makeatother

\renewcommand{\thefootnote}{\fnsymbol{footnote}}
\renewcommand\footnoterule{\vspace*{1pt}%
\color{cream}\hrule width 3.5in height 0.4pt \color{black}\vspace*{5pt}} 
\setcounter{secnumdepth}{5}

\makeatletter 
\renewcommand\@biblabel[1]{#1}            
\renewcommand\@makefntext[1]%
{\noindent\makebox[0pt][r]{\@thefnmark\,}#1}
\makeatother 
\renewcommand{\figurename}{\small{Fig.}~}
\sectionfont{\sffamily\Large\bf}
\subsectionfont{\normalsize}
\subsubsectionfont{\bf}
\setstretch{1.125} 
\setlength{\skip\footins}{0.8cm}
\setlength{\footnotesep}{0.25cm}
\setlength{\jot}{10pt}
\titlespacing*{\section}{0pt}{4pt}{4pt}
\titlespacing*{\subsection}{0pt}{15pt}{1pt}


\makeatletter 
\newlength{\figrulesep} 
\setlength{\figrulesep}{0.5\textfloatsep} 

\newcommand{\topfigrule}{\vspace*{-1pt}%
\noindent{\color{cream}\rule[-\figrulesep]{\columnwidth}{1.5pt}} }

\newcommand{\botfigrule}{\vspace*{-2pt}%
\noindent{\color{cream}\rule[\figrulesep]{\columnwidth}{1.5pt}} }

\newcommand{\dblfigrule}{\vspace*{-1pt}%
\noindent{\color{cream}\rule[-\figrulesep]{\textwidth}{1.5pt}} }
\newcommand{\psI}{phase~1}
\newcommand{\psII}{phase~2}
\newcommand{\rmd}{{\rm d}}
\newcommand{\dg}[1]{\textcolor{red}{#1}}
\makeatother


\twocolumn[
  \begin{@twocolumnfalse}

{\huge\textbf{Dynamics of switching processes: general results and
applications to intermittent active motion}}

\vspace{1em}
\sffamily
\begin{tabular}{m{4.5cm} p{13.5cm} }

& \noindent\large{Ion Santra\textit{$^{a\ddag}$}, Kristian St\o{}levik Olsen\textit{$^{b}$}, Deepak Gupta\textit{$^{c,d\ddag}$}} \\

&\noindent\normalsize{Systems switching between different dynamical phases is an ubiquitous phenomenon. The general understanding of such a process is limited. To this end, we present a general expression that captures fluctuations of a system exhibiting a switching mechanism. Specifically, we obtain an exact expression of the Laplace-transformed characteristic function of the particle's position. Then, 
the characteristic function is used to compute the effective diffusion coefficient of a system performing intermittent dynamics.  
Further, we employ two examples: 1) Generalized run-and-tumble active particle, and 2) an active particle switching its dynamics between generalized active run-and-tumble motion and passive Brownian motion. In each case, explicit computations of the spatial cumulants are presented. Our findings reveal that the particle's position probability density function
exhibit rich behaviours due to intermittent activity. Numerical simulations confirm our findings.

} \\

\end{tabular}

 \end{@twocolumnfalse} ]
 
 \vspace{0.6cm}


\renewcommand*\rmdefault{bch}\normalfont\upshape
\rmfamily
\section*{}
\vspace{-1cm}


\footnotetext{\textit{$^{a}$ Institute for Theoretical Physics, Georg-August Universit\"at G\"ottingen, 37077 G\"ottingen, Germany; E-mail: ion.santra@theorie.physik.uni-goettingen.de}}
\footnotetext{\textit{$^{b}$Institut für Theoretische Physik II - Weiche Materie, Heinrich-Heine-Universität Düsseldorf, D-40225 Düsseldorf, Germany}}
\footnotetext{\textit{$^{c}$~Department of Physics, Indian Institute of Technology Indore, Khandwa Road, Simrol, Indore-453552, India\\
$^d$Nordita, Royal Institute of Technology and Stockholm University,  Hannes Alfvéns väg 12, 23, SE-106 91 Stockholm, SwedenNordita, Royal Institute of Technology and Stockholm University, Roslagstullsbacken 23, SE-106 91 Stockholm, Sweden; Email: phydeepak.gupta@gmail.com}}



\footnotetext{\ddag~The authors contributed equally to this work.}


\section{Introduction}
\label{sec:intro}

Switching between different dynamical phases is a widespread phenomenon in a wide range of complex systems~\cite{yin2010hybrid,bressloff2020switching,tucci2022modeling}. Many examples can be found in a biological setting, including transitions between stages of the cell-cycle \cite{schafer1998cell,rombouts2021dynamic}, changes in animal motility patterns \cite{morales2004extracting, michelot2019state}, diffusion of particles in mammalian cells \cite{sabri2020elucidating,balcerek2023modelling}, and target-binding proteins that switch between sliding motion along DNA strands and three--dimensional excursions \cite{mirny2009protein,loverdo2009quantifying}.

Living or active systems are ideal candidates for displaying dynamical switching~\cite{buhl2006disorder,ariel2014individual,barone2023experimental,olsen2022collective}. For example, such systems are often capable of processing environmental information, leading to subsequent adaptation and changes in motility patterns  \cite{tkavcik2016information,caraglio2024learning,nasiri2023optimal}, a property not shared with their passive counterparts. 
The study of physical properties of life-like systems often falls
under the umbrella of active matter~\cite{ramaswamy2010mechanics, marchetti2013hydrodynamics,elgeti2015physics,bechinger2016active}. A plethora of examples can be found, ranging from motile cells \cite{berg1975chemotaxis} and synthetic active colloids \cite{ebbens2010pursuit} on the microscale all the way to macro-scale living organisms \cite{helbing2000simulating} and granular systems \cite{caprini2024emergent,aranson2007swirling,kumar2014flocking}.
Several theoretical investigations of active particles that undergo dynamical switching has been seen recently, including switching chirality~\cite{santra2021active,santra2021direction,olsen2021diffusion,das2023chirality} and particles that switch between many possible self-propulsion states ~\cite{krasky2018diffusion,frydel2022run,demaerel2018active}. Further, recent theoretical investigations show how such intermittent active phases can aid in spatial exploration and target search~\cite{peruani2023active, caraglio2024learning}.

\begin{figure}[t!]
\centering\includegraphics[width=0.9\hsize]{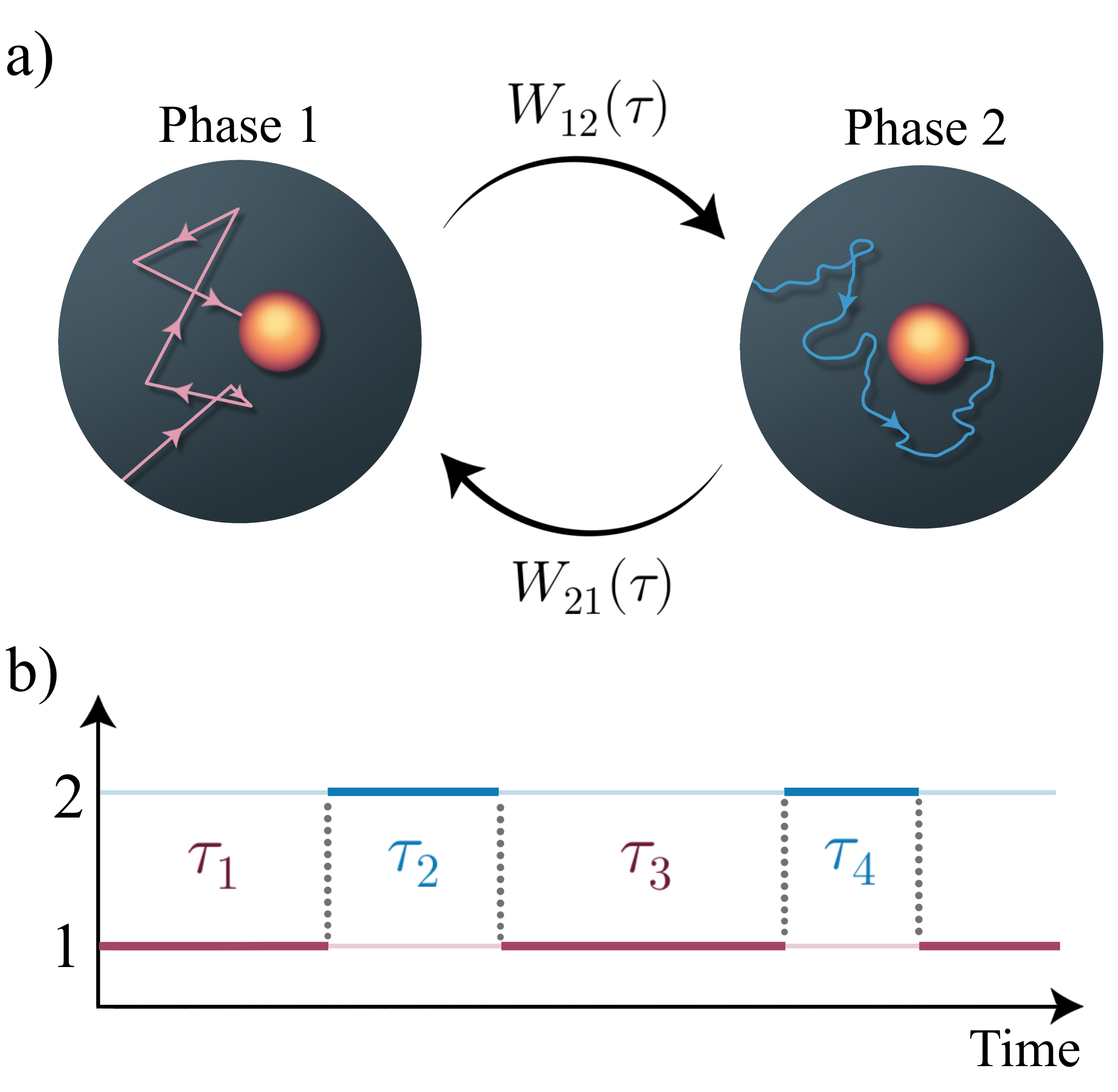}
\caption{a) Schematic representation of a two state stochastic process, where a particle iteratively switched between two arbitrary abstract dynamical phases. b) The particle switches between phase 1 and 
2 iteratively, with corresponding waiting-times $\tau$, where $\tau$ is chosen from the probability density functions $W_{1\to 2}(\tau)\equiv W_{12}(\tau)$ and $W_{2\to 1}(\tau)\equiv W_{21}(\tau)$.}
\label{f:scheme}
\end{figure}

A widespread type of switching found both on microscopic and macroscopic scales is activity that only acts intermittently, where directed motion is interspersed with passive or immobile phases \cite{perez2019bacteria,arcizet2008temporal,caraglio2023learning,maiuri2015actin,breoni2022one,bazazi2012intermittent,stojan2010functions,bartumeus2009behavioral,kramer2001behavioral,wilson2010boldness,antonov2024inertial,datta2024random,jamali2020active,hahn2020super,trouilloud2004head,higham2011climb,hafner2016run,datta2024intermittentrunmotilitybacteria}.
Interestingly, from an experimental perspective, intermittent activity can easily be achieved in systems of light-activated particles~\cite{buttinoni2012active}. Such experimental modelling of active behaviours has opened a door to design self-propelled particles with a broad range of novel behaviours~\cite{palacci2013living,buttinoni2013dynamical}, including the ability to make particles reverse their direction of motion with a specified asymmetry in the forward and backward speeds \cite{vutukuri2020light}.


\textcolor{black}{ In this paper, we present a general framework that captures the dynamics of a two-phase switching process 
(see Fig.~\ref{f:scheme}). In particular, we derive an exact expression for the (Fourier-Laplace transformed) probability density function of an observable that evolves according to the rules of two different phases. We employ the (Fourier-Laplace transformed) probability density function to compute the expression for the effective diffusion of the switching process, finding conditions on the waiting-time distribution that minimises the effective diffusivity. Motivated by the experiment based on light-activated particles~\cite{buttinoni2012active}, we study a minimal stochastic model in one-dimension involving a generalized run-and-tumble particle with intermittent passive phases. Using the results of the switching formalism, we first study the generalized run-and-tumble particle and characterise its position fluctuations, and find an optimal switching rate that maximises the long-time diffusion. Then, we explore the same process with intermittent passive phases by obtaining the first few position's 
cumulants exactly. We also discuss the time evolution of the position distribution of the active particle with intermediate passive phases, and 
show interesting 
short- (obtained by a perturbative analysis) and long-time behaviors.}

The rest of the paper is organised as follows. Section~\ref{sec:general} builds the general framework of the switching process and discusses the method to compute the fluctuations of the particle's position. Section~\ref{sec:deff} contains the derivation of the effective diffusion constant for a general scenario. In Sec.~\ref{sec:example}, we discuss two examples of switching processes, with detailed analysis of their time-dependent properties. Then, we discuss the position probability density function for a generalized run-and-tumble particle switching its behaviour to Brownian motion intermittently in Sec.~\ref{sec:dist}. Finally, we summarize the paper in Sec.~\ref{sec:summ}. Some of the detailed expressions are relegated to Appendices.

\section{General formalism}
\label{sec:general}
In the past several decades, the theoretical statistical physics community has explored the fascinating nature of dynamical switching processes by investigating its 
their
statistical properties.
Examples include modelling of dynamical coexistence of slow- and fast-diffusion in liquids and caging effects in glassy materials~\cite{singwi1960diffusive,chaudhuri2007universal}, as well as  
theoretical work on the dynamical disorder~\cite{zwanzig1989diffusion,zwanzig1992dynamical,zwanzig1990rate}. Additionally,  
stochastic resetting
~\cite{evans2020stochastic,mercado2020intermittent, santra2021brownian,gupta2020stochastic,mercado2022reducing, olsen_thermodynamic,olsen2024thermodynamic,frydel2024statistical}, diffusive systems with a switching diffusion coefficient~\cite{bressloff2017stochastic,baran2013feynman,grebenkov2019time,grebenkov2019unifying,goswami2022motion}, alternating dynamical phases \cite{benichou2011intermittent, bressloff2013stochastic,rukolaine2019intermittent,rukolaine2018model,hidalgo2021cusp}, and even the blinking of quantum dots~\cite{sadegh20141,niemann2013fluctuations} fall under the umbrella of the stochastic switching process. Several frameworks have been proposed to deal with systems exhibiting dynamical switching, such as hybrid and composite stochastic processes~\cite{van1979composite,bressloff2017stochastic,bressloff2023truncated}, piece-wise deterministic Markov processes~\cite{davis1984piecewise} and hidden Markov models~\cite{cappe2005inference,van2019classifying,bechhoefer2015hidden}.  Here, we consider a general approach to model systems with switching mechanisms, which
enable us to reproduce many some of the above examples.

In this paper, we consider a general dynamical process, where a system  switches stochastically between two phases, 1 and 2 (Fig.~\ref{f:scheme}). Stochastic dynamics characterize the evolution of this switching mechanism~\cite{Kampen1992}. Without loss of generality, we assume that the particle's trajectory begins in the \psI~and remains in this phase for a time-interval $\tau_1$ before switching its phase to \psII. It executes the phase 2 dynamics for a time-interval $\tau_2$, and then, its phase switches back to \psI. This process iterates until the observation time $t = \sum_{j=1} \tau_j$. The waiting time-intervals $\tau_1,\tau_3,\dotsc$ ($\tau_2,\tau_4,\dotsc$), that correspond to a switch from phase 1 to phase 2 (phase 2 to phase 1), 
are drawn from the probability density function $W_{12}(\tau)\equiv W_{1\to 2}$ ($W_{21}(\tau)\equiv W_{2\to 1}$), see Fig.~\ref{f:scheme}. 
Then, the survival probabilities $S_1(t)$
or $S_2(t)$, 
respectively, denote the probability that the particle starting from 
\psI~or~\psII~has not switched to its complementary phase until time $t$. Mathematically, these read as
\begin{subequations}
\begin{align}
S_1(t)\equiv\int_t^\infty \rmd \tau~W_{12}(\tau)\ ,\\
S_2(t)\equiv\int_t^\infty \rmd \tau~W_{21}(\tau)\ .
\end{align}
\end{subequations}

To proceed further, we write the propagator of the particle (probability density function of particle's position) evolving solely in  
\psI~(or 
phase 2) until time $t$, having started from position $\vec x_0$ at time $t_0$; this is denoted by $P_{\rm 1}(\vec x,t|\vec x_0,t_0)$ [or $P_{\rm 2}(\vec x,t|\vec x_0,t_0)$]{\color{black}, where $\vec x$ is the particle's position in any spatial dimension.}. Further, we assume no potential energy landscapes in either of these phases; therefore, the particle's dynamics in each phase depends on the distance covered from its initial location in time $t-t_0$. 
Thus, we write
\begin{align}
P_{j}(\vec x-\vec x_0,t-t_0)\equiv P_{j}(\vec x-\vec x_0,t-t_0|0,0)
= P_{j}(\vec x,t|\vec x_0,t_0)\ ,
\label{eq:isotropic}
\end{align}
for the phase index $j=\{{\rm 1, 2}\}$.

For simplicity of notation, we assume that the particle starts from the origin ($\vec x=(0,0)$) at time $t=0$ in phase 1. Let the time-interval between the $i$'th and the $(i+1)$'th switching event be $\tau_{i+1}$, and the displacement in that time-interval be $\vec x_{i+1}$.   Then, the propagator accounts the contributions arising from trajectories experiencing different number of switching events until time $t$.

Let us first compute the contribution to the probability density function of the particle's position $\vec x$ in time $t$, from a trajectory that has not experienced a single switching event until time $t$. This is given by
\begin{align}\label{eq:p0}
\phi_0(\vec x,t) \equiv S_1(t)~P_{1}(\vec x,t)\ .
\end{align}
Note that, the right-hand side of the above equation~\eqref{eq:p0} is in a product form. This is because the switching between the phases and the individual dynamics in each phase are independent events.

Now, suppose a single switching event occurs from phase 1 to 2 in time-interval $\tau_1$, and in the remaining time $\tau_2 = t-\tau_1$, there is no switching occurs. Then, we have
\begin{align}
\phi_1(\vec x,t)&\equiv \int_{-\infty}^{\infty}~\rmd \vec x_1\int_{-\infty}^{\infty}~\rmd \vec x_2~\int_0^t \rmd\tau_1~W_{12}(\tau_1)~P_1(\vec x_1,\tau_1)\nonumber\\&\times \int_0^t \rmd\tau_2~ S_2(\tau_2)~P_{2}(\vec x_2,\tau_2)\delta(\vec x-\vec x_1-\vec x_2)\delta(t-\tau_1-\tau_2)\ .
\end{align}
Several comments are in order. The first part, i.e., $W_{12}(\tau_1)P_1(\vec x_1,\tau_1)$, indicates that the particle is evolved by the phase 1 dynamics for a time-interval $\tau_1$, and then, underwent a switching event. Further, the second part, i.e., $S_{2}(\tau_2)P_2(\vec x_2,\tau_2)$, is for the particle that followed the phase 2 dynamics and did not undergo further switching events in the remaining time-interval $\tau_2$. The intervals $\tau_1$ and $\tau_2$ are arbitrary and hence they have to be integrated over along with a delta function ensuring that the total observation time is $t = \tau_1 + \tau_2$. Another delta function ensures that the particle reaches $\vec x$ after travelling $\vec x_1$ and $\vec x_2$, respectively, in phase 1 and 2.

The upper limits of the $\tau_i$ integrals can be set to $+\infty$, since the delta-function still ensures that the region of integration in the $(\tau_1,\tau_2)$ plane remains the same. Then, we have
\begin{align}
\label{zero}
\phi_1(\vec x,t)&\equiv \int_{-\infty}^{\infty} \rmd \vec x_1~\int_{-\infty}^{\infty}~\rmd \vec x_2~\int_0^\infty \rmd\tau_1~W_{12}(\tau_1)P_1(\vec x_1,\tau_1)\nonumber\\
&\times \int_0^\infty \rmd\tau_2~ S_2(\tau_2)P_{2}(\vec x_2,\tau_2)\delta(\vec x-\vec x_1-\vec x_2)\delta(t-\tau_1-\tau_2)\ .
\end{align}

It is easy to generalize the above procedure for any number of switching events. For instance, for an odd number of switching events (particle ends up in phase 2 at time $t$), the contribution to the probability density of particle's position is
\begin{align}
\label{odd}
\phi_{2m+1}(\vec x,t)&\equiv\prod_{j=1}^{2m+2}\int_{-\infty}^{\infty}d\vec x_j\prod_{k=1}^{2m+2}\int_{0}^{\infty} d\tau_k~W_{12}(\tau_1)P_{1}(\vec x_1,\tau_1)\cr
&\times W_{21}(\tau_2)P_{2}(\vec x_2,\tau_2)\dotsc W_{12}(\tau_{2i+1})P_{1}(\vec x_{2i+1},\tau_{{2i+1}})\cr
&\times W_{21}(\tau_{2i})P_{2}(\vec x_{2i},\tau_{2i}) \dotsc S_{2}(\tau_{2m+2})P_{2}(\vec x_{2m+2},\tau_{2m+2})\cr
&\times \delta\left(\vec x-\sum_{i=1}^{2m+2}\vec x_i\right)\delta\left(t-\sum_{i=1}^{2m+2}\tau_i\right)\ ,
\end{align}
for a positive integer $m\geq 0$. Similarly, for even number of switching events (particle ends up in phase 1 at time $t$) the probability density of particle's position becomes ($m\geq 0$)
\begin{align}
\label{even}
\phi_{2m+2}(\vec x,t)&\equiv \prod_{j=1}^{2m+3}\int_{-\infty}^{\infty}d\vec x_j\prod_{k=1}^{2m+3}\int_{0}^{\infty} d\tau_k~W_{12}(\tau_1)P_{1}(\vec x_1,\tau_1)\cr
&\times  W_{21}(\tau_2)P_{2}(\vec x_2,\tau_2)\dotsc W_{12}(\tau_{2i+1})P_{2}(\vec x_{2i+1},\tau_{{2i+1}})\cr
&\times W_{21}(\tau_{2i})P_{2}(\vec x_{2i},\tau_{2i}) \dotsc S_{1}(\tau_{2m+3})P_{1}(\vec x_{2m+3},\tau_{2m+3})\cr
&\times \delta\left(\vec x-\sum_{i=1}^{2m+3}\vec x_i\right)\delta\left(t-\sum_{i=1}^{2m+3}\tau_i\right)\ .
\end{align}

Therefore, using Eqs.~\eqref{zero},~\eqref{odd}, and~\eqref{even}  the probability density function particle's position $\vec x$ at time $t$ is 
\begin{align} 
p(\vec x,t)\equiv \sum_{n=0}^{\infty}\phi_{n}(\vec x,t)=\sum_{m=0}^{\infty}\left[\phi_{2m}(\vec x,t)+\phi_{2m+1}(\vec x,t)\right]\ .
\label{eq:ptot}
\end{align}
The summation on the right-hand side~\eqref{eq:ptot} can be evaluated in the Fourier-Laplace transform defined by,
\begin{align}
\tilde{\bar{f}}(\vec k,s)=\mathcal{L}[\bar f(k,t) ]=\int_{0}^{\infty}\rmd t~e^{-st}\int_{-\infty}^{\infty}\rmd \vec x~e^{i\vec k\cdot\vec x} f(\vec x,t)\ ,
\end{align}
where $s$ and $\vec k$, respectively, are the conjugate variables with respect to time $t$ and position $\vec x$, and the overhead bar and tilde, respectively, denote the Fourier and Laplace transformed expressions. 
This gives 
\begin{align}
\tilde{\bar{p}}(\vec k,s)=\sum_{m=0}^{\infty}[\tilde{\bar{\phi}}_{2m}(\vec k,s)+\tilde{\bar{\phi}}_{2m+1}(\vec k,s)]\ , \label{phi-eqn}
\end{align}
for
\begin{align}
\tilde{\bar{\phi}}_{2m}(\vec k,s)=&\big(\mathcal{L}[W_{12}(t)\bar{P}_1(\vec k,t)]\big)^m\big(\mathcal{L}[W_{21}(t)\bar{P}_{2}(\vec k,t)]\big)^m\nonumber\\
&\times \mathcal{L}[S_{1}(t)\bar{P}_{1}(\vec k,t)]\ ,\label{eq:p-ev}\\
\tilde{\bar{\phi}}_{2m+1}(\vec k,s)=&\big(\mathcal{L}[W_{12}(t)\bar{P}_1(\vec k,t)]\big)^{m+1}\big(\mathcal{L}[W_{21}(t)\bar{P}_{2}(\vec k,t)]\big)^m\nonumber\\
&\times\mathcal{L}[S_{2}(t)\bar{P}_{2}(\vec k,t)]\ . \label{eq:p-odd}
\end{align}
Then, we perform the summation in Eq.~\eqref{phi-eqn}, and this gives Fourier-Laplace transformed probability density function:
\begin{align}
\tilde{\bar{p}}(\vec k,s)=\dfrac{\mathcal{L}[S_1(t)\bar P_1(\vec k,t)]+\mathcal{L}[W_{12}(t)\bar P_1(\vec k,t)]\mathcal{L}[S_2(t)\bar P_{2}(\vec k,t)]}{1-\mathcal{L}[W_{12}(t)\bar{P}_1(\vec k,t)]\mathcal{L}[W_{21}(t)\bar{P}_{2}(\vec k,t)]}\ .\label{eq:fullgeneral}
\end{align}
The above equation~\eqref{eq:fullgeneral} is the characteristic function for the switching process in the Laplace space. 
The above formula is true for both Markovian and non-Markovian switching protocols. Similar results are found in \textcolor{black} {Refs.~\cite{singwi1960diffusive,detcheverry2015non,thiel2012anomalous,angelani2013averaged, dieball2022scattering}.}

\section{Effective diffusion coefficients} 
\label{sec:deff}
The propagator~\eqref{eq:fullgeneral} is applicable for a wide range of switching processes. However, 
in several 
situations, it might be difficult to analytically invert the Fourier-Laplace transformed probability distribution~\eqref{eq:fullgeneral}. 
Nevertheless, 
the propagator~\eqref{eq:fullgeneral} is useful to extract dynamical information for a general class of systems. Herein, we derive an exact expression for the effective diffusion coefficient valid for any two unbiased dynamical phases.

The $n$'th order position fluctuations along the one dimensional coordinate $x$ can be obtained by differentiating the Laplace transformed characteristic function $\tilde{\bar{p}}(\vec k,s)$~\eqref{eq:fullgeneral} with respect to $k_x$ and setting $k_i=0$ (for all $i=x,y,z,\dotsc$), and then, 
inverting the Laplace transform. For simplicity, we consider a one dimensional processes from now on. The $n$th order position moment is 
\begin{align}
\la x^n(t) \ra=(-i)^n\mathcal{L}^{-1}\left[\frac{\partial^n \tilde{\bar{p}}(k,s)}{\partial k^n}\Bigg|_{k=0}\right].
\label{momentsgen}
\end{align}

The second derivative of the switching propagator~\eqref{eq:fullgeneral} takes the form 
\begin{align}
\label{eq:twoderivatives}
    \tilde{\overline{p}}''(k=0,s)  &= \frac{\mathcal{A}(s)}{1 -\tilde W_{12}(s) \tilde W_{21}(s)}  + \frac{\mathcal{B}(s)}{[1 -\tilde W_{12}(s) \tilde W_{21}(s)]^2}\ ,
\end{align}
where $prime$ indicates a derivative with respect to $k$, and we used the zero-mean property $\tilde P_i'(k=0,t) =0$ as well as normalization $\tilde P_i(k=0,t) =1$ for both phases. For brevity, we have introduced the functions
\begin{align}
    \mathcal{A}(s) &\equiv \mathcal{L}[S_1(t) P_1''(k,t)] +\tilde{W}_{12}(s) \mathcal{L}[S_2(t) P_2''(k,t)] \nonumber \\
    &+ \tilde{S}_2(s) \mathcal{L}[W_{12}(t) P_1''(k,t)]  \bigg |_{k=0} \ , \\
        \mathcal{B}(s) &\equiv \big[ \tilde S_1(s) + \tilde S_2(s) \tilde W_{12}(s) \big]  \big[\tilde W_{21}(s) \mathcal{L}[W_{12}(t) P_1''(k,t)] \nonumber \\
        &+\tilde W_{12} \mathcal{L}[W_{21}(t) P_2''(k,t)] \big]\bigg |_{k=0} \label{eq:calB} \ .
\end{align}

To obtain the long-time mean squared displacement and the effective diffusion coefficient, we have to consider the dominant pole in $s$ as $s\to 0$. \textcolor{black}{First, we note that the term $[1 - \tilde W_{12}(s)\tilde W_{21}(s)]$ in the denominator of both 
fractions in~Eq.~\eqref{eq:twoderivatives} can be written for small $s$-values as\footnote{{\color{black}$\tilde W_{ij}(s)=\int_0^t~dt~e^{-s t}W_{ij}(t) \approx 1 - s \int_0^t~dt~t~W_{ij}(t)+\mathcal{O}(s^2)  =1 - s \langle t \rangle +\mathcal{O}(s^2).$}}
\begin{align}
  1 - \tilde W_{12}(s)\tilde W_{21}(s) = s (\langle \tau_1\rangle+\langle \tau_2\rangle)+\mathcal{O}(s^2)\ ,   
\end{align}
by noting that $W_{ij}(t)$ denotes the  probability density functions and 
 $\tilde W_{ij}(s=0)=1$.}
Since the second term in Eq.~\eqref{eq:twoderivatives} has denominator $[1 - \tilde W_{12}(s)\tilde W_{21}(s) ]^2$, this term will be the leading contribution at late times (e.g. small $s$).
Hence, the 
long-time behaviour is given by
\begin{equation}
    \widetilde {\langle x^2\rangle} = - \frac{\mathcal{B}(0)}{[s (\langle \tau_1\rangle+\langle \tau_2\rangle)]^2} + \mathcal{O}(s^{-1})\ .
\end{equation}
 where $\mathcal{B}(0)\leq 0$. \textcolor{black}{Upon inverting the Laplace transform, one finds that at long times, $\la x^2(t)\ra\approx 2D_\text{eff} t$, with the effective diffusion coefficient:} 
\begin{equation}
    D_\text{eff} \equiv \lim_{t\to \infty} \frac{\langle x^2(t) \rangle}{2 t} = - \frac{\mathcal{B}(0)}{ 2(\langle \tau_1\rangle+\langle \tau_2\rangle)^2} \ .
\end{equation}
Finally, from Eq.~\eqref{eq:calB} we can calculate $\mathcal{B}(0)$ by using the fact that 
$\tilde S_i(0) = \langle \tau_i \rangle$, $\tilde W_{ij}(0) = 1$,  as well as
\begin{equation}
    \mathcal{L}[W_{ij}(t) P_i''(k=0,t)](s=0)  =- \int_0^\infty dt~W_{ij}(t) \langle x^2(t)\rangle_i 
\end{equation}
By combining the above we find the effective diffusion coefficient 
\begin{equation}\label{eq:deffgeneral}
    D_\text{eff} = \frac{\int_0^\infty dt~W_{12}(t) \langle x^2(t)\rangle_1 }{2(\langle \tau_1\rangle+\langle \tau_2\rangle)} + \frac{\int_0^\infty dt~W_{21}(t) \langle x^2(t)\rangle_2 }{2(\langle \tau_1\rangle+\langle \tau_2\rangle)}\ .
\end{equation}
 This expression is valid for any two symmetric processes with mean squared displacements $\langle x^2(t)\rangle_i$ ($i=1,2$), for any switching protocol as long at the mean waiting times are well-defined $\langle \tau_i\rangle < \infty$. When the particle is immobile in one of the phases 
 , e.g., $\langle x^2(t) \rangle_2 = 0$, and the waiting time in the first phase is exponential, we recover the results obtained in Ref.~\cite{Olsen_2024}. When both phases are diffusive and the waiting times exponential, we recover existing results from models with switching diffusion coefficients \cite{grebenkov2019time}.

\begin{figure}[t!]
    \centering
    \includegraphics[width = 8.5cm]{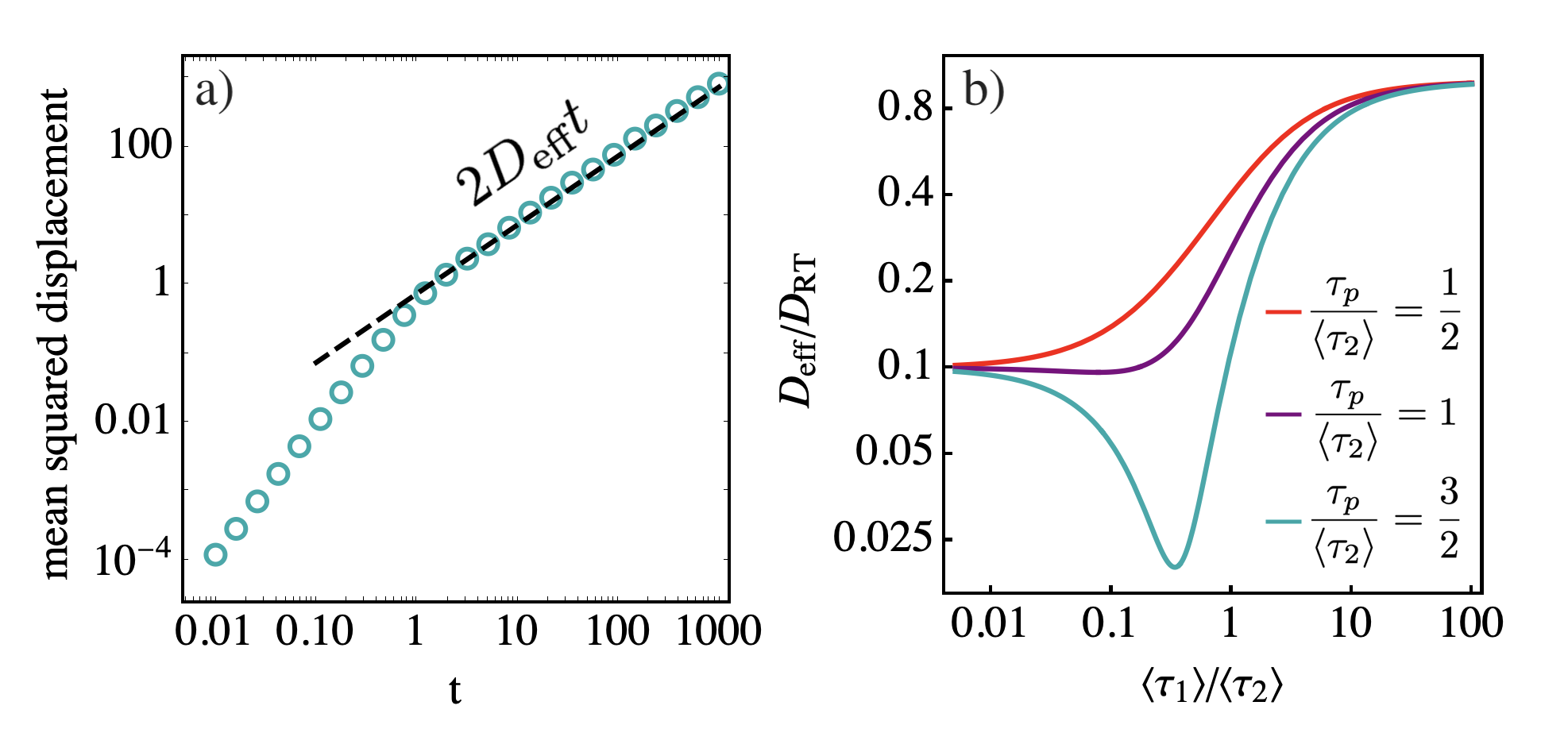}
    \caption{a) Mean squared displacement of an active run-and-tumble particle intermittently switching to a passive phase. Active durations are drawn from a gamma distribution~\eqref{eq:gamma} with shape index $\gamma = 3$. The dashed line shows theoretical prediction based on the effective diffusion coefficient [i.e., Eq.~\eqref{eq:deff_active_laplace} with \eqref{gm-lpt}], while circles show numerical simulation data. For simplicity, the passive phases have deterministic duration $\langle \tau_2 \rangle = 1/2$. Remaining parameters used are $v=1,~D=1,~\tau_{\rm p} = 0.5$. b) Effective diffusivity rescaled with $D_\text{RT}\equiv v^2 \tau_{\rm p}$ as a function of the ratio of active durations to passive durations, again for the case when active durations drawn from a gamma distribution~\eqref{eq:gamma} with shape index $\gamma = 3$. In this case, we have set $D/D_\text{RT} = 0.1$, so that passive phases are slow compared to the active motion.  }
    \label{fig:gamma}
\end{figure}

 In the case of intermittent active systems where the motion is interspersed with passive periods, we can use the above general formula~\eqref{eq:deffgeneral} together with the known expression for the mean squared displacement of an active particle to obtain the effective diffusion coefficient. We let phase $1$ correspond to the active phase, and phase $2$ to the passive Brownian motion. Active particles are characterized by a persistence time $\tau_{\rm p}$, and the mean squared displacement takes the form \cite{romanczuk2012active,santra2020run}
 \begin{equation}\label{eq:active_phase_msd}
     \langle x^2(t)\rangle_1 = 2 v^2 \tau_{\rm p} \left [ t +\tau_{\rm p} \left(e^{-t/\tau_{\rm p}}-1\right)  \right]\ ,
 \end{equation}
 where $v$ is the self-propulsion speed.
 As the persistence time becomes smaller, a passive-like behaviour appears. Hence, active-passive switching can be thought of as the motion of an active particle that switched from persistent motion, to less persistent motion, for example as part of an intermittent search strategy. 
 
The passive phase is diffusive with $ \langle x^2(t)\rangle_2 = 2 D t $ where the passive diffusion coefficient is $D$. Eq. (\ref{eq:deffgeneral}) then gives 
\begin{align}
\label{eq:deff_active_laplace}
     D_\text{eff} =& \frac{\langle \tau_1\rangle }{\langle \tau_1\rangle+\langle \tau_2\rangle}v^2 \tau_{\rm p}  +  \frac{ \langle \tau_2\rangle}{\langle \tau_1\rangle+\langle \tau_2\rangle} D \nonumber \\
     & + \frac{v^2 \tau_{\rm p}^2 \left(\tilde W_{12}(\tau_{\rm p}^{-1}) - 1\right) }{\langle \tau_1\rangle+\langle \tau_2\rangle}\ .
\end{align}
{\color{black} We note here that while the long-time effective diffusion coefficient of the active particle without passive phases would have been simply $v^2 \tau_p$, the long-time diffusion in the presence of switching also depends on the transient behavior coming from the mean squared displacement of the active phase, Eq. (\ref{eq:active_phase_msd}. While this transient is irrelevant at late times in the absence of switching, the recurrent restarts of the active phase makes it relevant at late times once switching is considered.} 
The above calculation remains general, and is valid for any waiting time density $W_{12}(t)$ in the active phase.

Therefore, 
a relevant question is how the choice of distribution of the active phase durations affect the effective diffusion coefficient.  Since 
$\tilde W_{12}(\tau_{\rm p}^{-1}) = \langle e^{-\tau/\tau_{\rm p}}\rangle$, we note that by Jensen's inequality, we have $ \tilde W_{12}(\tau_{\rm p}^{-1}) \geq e^{-\langle \tau_1 \rangle/\tau_{\rm p}}$. At fixed mean durations of the active phase, this inequality is saturated only by the deterministic case $W_{12}(\tau) = \delta(\tau-\langle\tau_1\rangle)$. Hence, at fixed durations of the two phases, the effective diffusivity is always smallest when the duration in the active phase is deterministic:
\begin{align}
    D_\text{eff}[W] \geq D_\text{eff}[\delta] \equiv  D_\text{eff}^\text{min}\ . \label{left-bd}
\end{align}
Hence, fluctuations in the active phase durations are needed for effective spatial exploration.

Furthermore, since $\tilde W_{12}(s) \leq 1$ by monotonicity of integrals, we also have the upper bound
\begin{equation}
    D_\text{eff}^\text{max} = \frac{\langle \tau_1\rangle }{\langle \tau_1\rangle+\langle \tau_2\rangle}v^2 \tau_{\rm p}  +  \frac{ \langle \tau_2\rangle}{\langle \tau_1\rangle+\langle \tau_2\rangle} D \ .\label{right-bd}
\end{equation}
Hence, the effective diffusion coefficient is bounded as $D_\text{eff}^\text{min} \leq D_\text{eff} \leq D_\text{eff}^\text{max}$, with the explicit expressions given above~\eqref{left-bd} and \eqref{right-bd}.

To illustrate the above results, we  specifically consider the case when the active durations are drawn from a gamma distribution:
\begin{equation}
\label{eq:gamma}
    W_{12}(\tau) = \frac{\tau^{\gamma-1}}{\tau_0^\gamma~\Gamma(\gamma)}e^{-\tau/\tau_0}\ ,
\end{equation}
where $\Gamma(\cdot)$ is the Gamma-function. Its Laplace transform reads
\begin{equation}
    \tilde W_{12}(\tau_{\rm p}^{-1}) = \left(\frac{\tau_{\rm p}+\tau_0}{\tau_{\rm p}}\right)^{-\gamma}\ , \label{gm-lpt}
\end{equation}
From this, the effective diffusion coefficient can be easily obtained from Eq.~\eqref{eq:deff_active_laplace}. We note that in this case, the dependence on the duration of the passive Brownian phase enters only through its first moment, and is insensitive to further details of the distribution $W_{21}(\tau)$.

Figure \ref{fig:gamma}a shows the mean squared displacement obtained numerically together with the theoretical prediction [i.e., Eq.~\eqref{eq:deff_active_laplace} with \eqref{gm-lpt}], verifying that the system's long-time behaviour will be diffusive with the effective diffusion coefficient $D_{\rm eff}$~\eqref{eq:deffgeneral}.

Figure \ref{fig:gamma}b shows the effective diffusivity rescaled with the late-time pure run-and-tumble diffusion coefficient $D_\text{RT} \equiv v^2 \tau_{\rm p}$, as a function of the ratio $\langle \tau_1\rangle/\langle \tau_2\rangle$.

\section{Model: active particles with intermittent passive phases}
\label{sec:example}
\begin{figure}
    \centering
    \includegraphics[width=0.75\hsize]{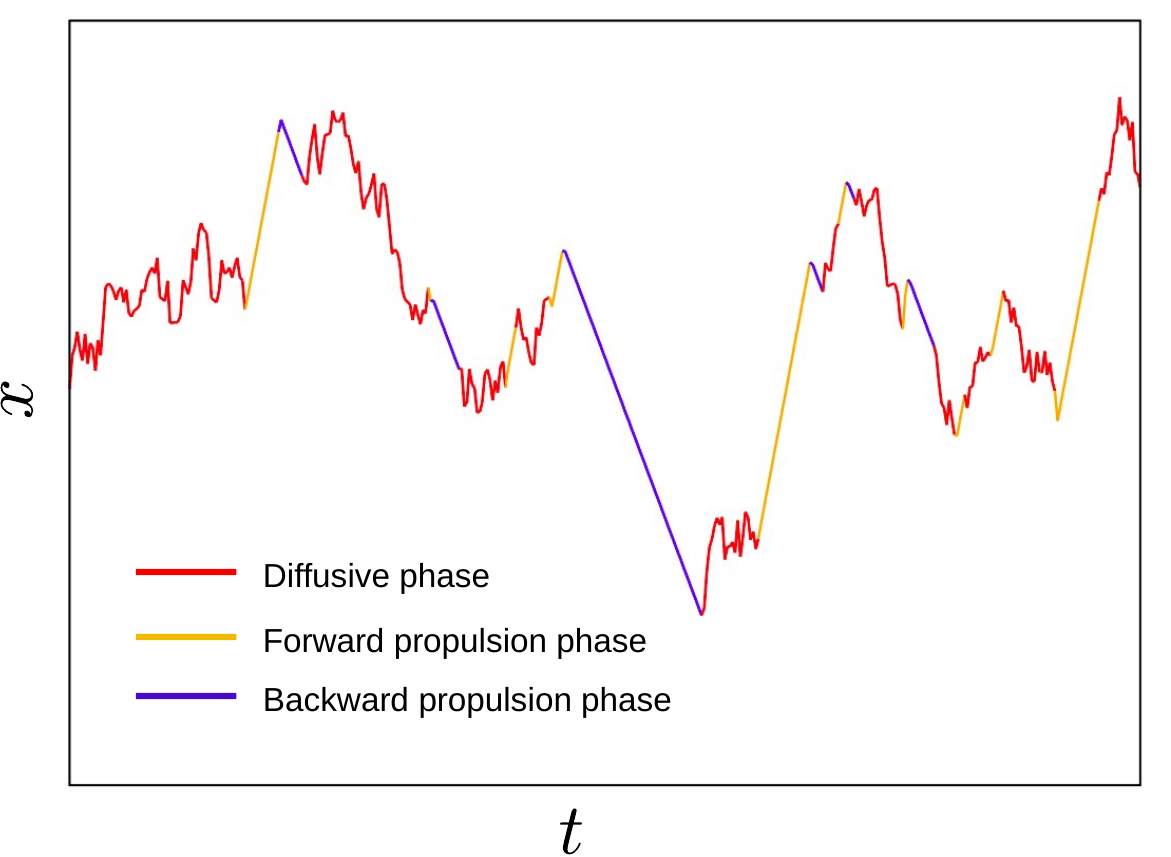}
    \caption{
    Trajectory of a generalized run-and-tumble particle (orange and blue) with intermittent passive phases (red).} 
    \label{fig:traj}
\end{figure}
Section~\ref{sec:general} discusses
the framework 
of modelling processes switching between two arbitrary homogeneous dynamical phases. While general properties of such systems can be studied for symmetric processes, such as the effective diffusion coefficients in section~\ref{sec:deff}, the framework is also capable of handling more involved processes with bias. In this section, we consider in detail the case of a (potentially biased) run and tumble particle (RTP) that intermittently switches to passive Brownian motion {\color{black} (see Fig. \ref{fig:traj})}.

Let us consider the dynamics of an overdamped self-propelled particle in one dimension.
Similar to the dynamics of the light-activated Janus particles~\cite{vutukuri2020light}, 
this particle has three phases of motion: (i) A forward propulsion with a speed $v_{\rm f}$, when irradiated by a UV light; (ii) A backward propulsion with a speed $v_{\rm b}$ when irradiated by a green light; and (iii) A diffusive motion 
in the absence of light. Mathematically, 
the following equations model these phases:
\begin{align}
\dot x(t)=\begin{cases}v_{\rm f}\quad &\text{forward propulsion},\\
-v_{\rm b}\quad &\text{backward propulsion},\\
\sqrt{2D}\eta(t)&\text{passive diffusion},
\end{cases} 
\label{eq:gmodel0}
\end{align}
where $v_{\text{b,f}}\geq 0 $, and $\eta(t)$ is a delta-correlated Gaussian white noise with zero mean: $\la\eta(t)\eta(t')\ra=\delta(t-t')$ and 
$D$ is the diffusion coefficient for the phase (iii).

We define the following switching mechanism between the three phases. The switching between the forward to backward propulsion and backward to forward propulsion (i.e., switching between UV and green light, and vice versa, respectively) happens at a rate $\alpha_{\text{fb}}$ and $\alpha_{\text{bf}}$, respectively. The switching from the \emph{active} propulsion phase to the \emph{passive} diffusive phase, (i.e., any light being turned off) occurs at a rate $\bar r_{\text{ap}}$; while the switching from the \emph{passive} to the \emph{active} phase, (i.e., any light being turned on) occurs at a rate $\bar r_{\text{pa}}$. 
Notice that we assume that when the lights are turned on, the particle can be either exposed to the UV or green light with equal probability. Furthermore, for simplicity, we consider a case where the particle always starts from the origin $x=0$ at time $t=0$ from any one of the active phases with equal probability.

It is always convenient to define a characteristic length- $\ell\equiv v_{\rm f}/\alpha_{\text{fb}}$ and a time-scale $\mathcal{T}\equiv \ell/v_{\rm f}$.
{\color{black}Then, Eq.~\eqref{eq:gmodel0} can be expressed in the dimensionless variables $x' = x/\ell$ and $t' = t/\mathcal{T}$, and dimensionless (rescaled) diffusion coefficient $D'\equiv D \mathcal{T}/\ell^2$.}  For convenience, we drop prime from these 
{\color{black}quantities}, and write
\begin{align}
\dot x(t)=
\begin{cases}
1\quad &\text{forward propulsion}, \\
-\theta\quad &\text{backward propulsion}\\
\sqrt{2D}\eta(t)&\text{passive diffusion}
\end{cases}\ ,
\label{eq:gmodel2}
\end{align}
where 
$\theta\equiv v_{\rm b}/v_{\rm f}$ measures the asymmetry in the backward to forward speed. The noise is again delta-correlated in time $\langle \eta(t_1)\eta(t_2)\rangle=\delta(t_1-t_2)$. In the dimensionless units, the relative switching between backward to forward motion occurs at a rate $A=\alpha_{\rm bf}/\alpha_{\rm fb}$, and active-passive switching rates become $r_{\text{ap}}$ and  $r_{\text{pa}}$.

In the following sections, we investigate this three-phase dynamics using the general switching formalism. To this end, we divide the dynamics Eq.~\eqref{eq:gmodel2} into (1) an active phase where the particle switches between a forward propulsion state and a backward propulsion state; and (2) a passive diffusive phase. The active phase, a generalized version of the one-dimensional RTP, is itself an example of a switching process, which we first analyze in the following subsection.

\subsection{Generalized run-and-tumble particles in one dimension}\label{genrtp}
{\color{black} Several interesting extensions of the symmetric run-and-tumble particle has been considered over the years, including non-uniform and anisotropic tumbles \cite{villa2020run,sevilla2020two,loewe2024anisotropic}. Here, we apply the above framework to the case of a run-and-tumble process in one dimension with biased run speeds \cite{lopez2014asymmetric}, before subsequently including intermittent passive phases.} 

Here, we discuss the scenario (1). The forward and backward propulsion are given by Eq.~\eqref{eq:gmodel2}(top two lines), they have the respective propagators,
\begin{align}
    P_{\rm f}(x,t)&=\delta(x-t)\ ,\\
    P_{\rm b}(x,t)&=\delta(x-\theta t)\ ,
\end{align}
which in the Fourier-Laplace space read:
\begin{align}
\tilde{\bar{P}}_{\rm f}(k,s)&=\frac{1}{s-i k}\ ,\\
\tilde{\bar{P}}_{\rm b}(k,s)&=\frac{1}{s+i k \theta}\ .
\end{align}

Suppose the particle begins in the forward phase. Using Eq.~\eqref{eq:fullgeneral} for the exponential switching rate $A$ measuring relative rate with respect to forward-to-backward propulsion rate $\alpha_{\rm fb}$, we have for the Fourier-Laplace transform of the distribution \textcolor{black}{ $\tilde{\bar{p}}_\text{f}(k,s)$:
\begin{align}
\tilde{\bar{p}}_\text{f}(k,s)=\dfrac{1+s+A+i k\theta}{s+ik \theta+(s-ik)(s+A+ik\theta)}\ .
\end{align}
Similarly, for the particle starting in the backward phase,
\begin{align}
\tilde{\bar{p}}_\text{b}(k,s)=\dfrac{1+s+A-i k\theta}{s+ik \theta+(s-ik)(s+A + ik \theta)}\ .
\end{align}
}

We assume that the initial propulsion state of the active particle is chosen from the stationary state, \textcolor{black}{ $p_\text{f}=A/(1+A)$ and $p_\text{b}=1/(1+A)$.} Using this, the Fourier-Laplace transform of the position distribution becomes
\begin{align}
    \tilde{\bar{p}}(k,s)=&{\color{black}\tilde{\bar{p}}_\text{f}(k,s)p_\text{f}+\tilde{\bar{p}}_\text{b}(k,s)p_\text{b}}\cr
    =&\frac{k(1-A \theta)+i(1+A)^2+i\, s(1+A)}{(1+A)\left(s+ik \theta+(s-ik)(A+s+ik\theta)\right)}\ .
    \label{pks:active}
\end{align}

It is difficult to invert the above equation exactly to obtain a closed-form expression for the position distribution at all times. Nevertheless, $n$'th moment of the position fluctuations can be obtained by using Eq.~\eqref{momentsgen}. 
It turns out that the RTP is characterized by a non-zero mean,
\begin{align}
    \la x(t)\ra=\dfrac{A-\theta}{(1+A)}~t\ ,\label{g:rtp:mean}
\end{align}
as expected. Interestingly, the mean vanishes when $A=\theta$. We call this the `unbiased limit' of the generalized RTP, where it has a zero mean in spite of having different velocities in the forward and backward propulsion phases. More precisely, the particle spends less time in the propulsion state with higher velocity, and more time in the lower velocity state, such that there is no net drift. This 
can be intuitively understood as follows. The system is initialized from the stationary probabilities of its forward and backward propulsion states, thus, at any time $t$, the average-time spent by the particle in the forward and backward phase, respectively, is $A t/(1+A)$ and $t/(1+A)$, which amounts to  the total displacements in the respective phase $A t/(1+A)$ and $A\theta t/(1+A)$, respectively. The sum of these two displacements lead to \eqref{g:rtp:mean}. Interestingly, the mean vanishes for $A=\theta$. 
Notice that this is different from the `standard RTP' limit (i.e., $A=\theta=1$), where again the mean is zero. 
\begin{figure}
    \centering
\includegraphics[width=\hsize]{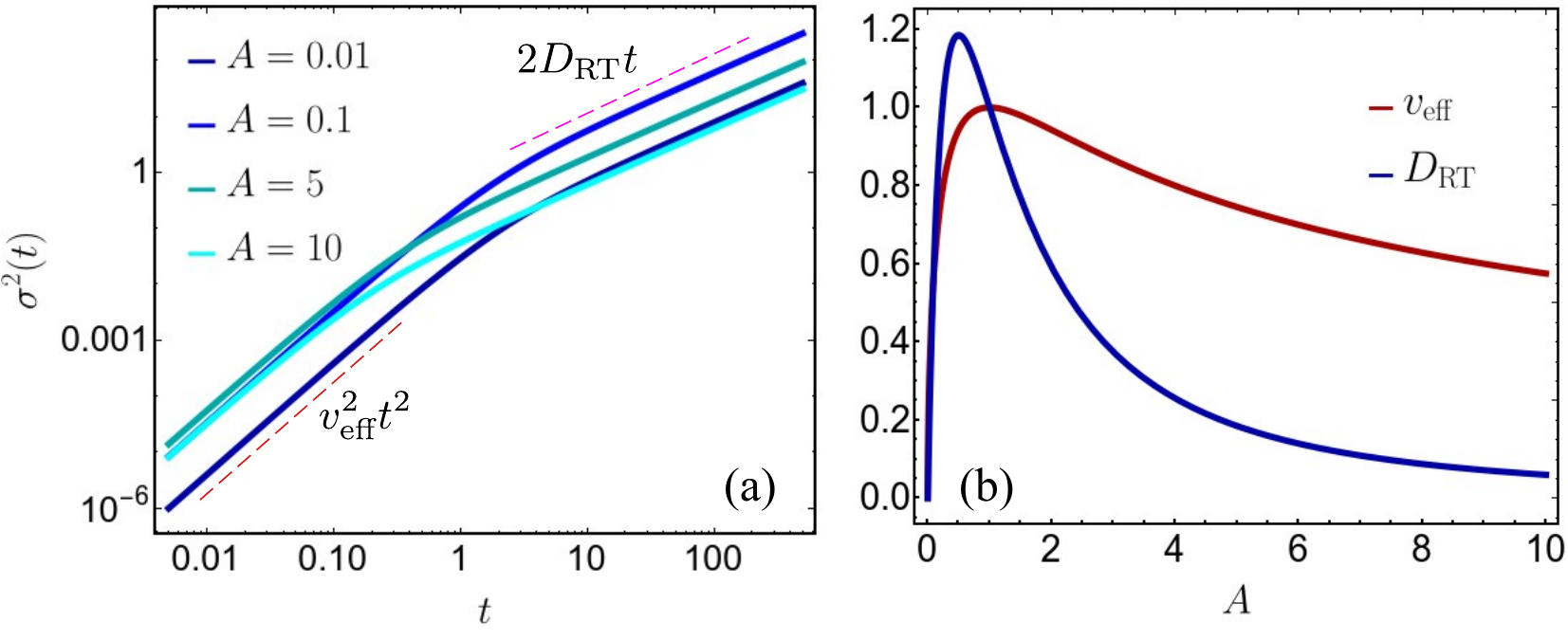}
    \caption{Position fluctuations of a generalized RTP. Panel (a) shows the variance of the particle for different values of $A$, which crosses over from an initial ballistic phase to a late time diffusive phase at times $t\sim \min (1,A^{-1})$. Color intensity increases with increasing the value of relative switching rate between backward to forward motion, $A$. Panel (b) shows the variation of the effective velocity $v_{\rm eff}$ of the fluctuations and the effective diffusivity $D_{\rm eff}$ with respect to $A.$ For both plots $\theta=1.$
    }
    \label{fig:rtp:msd}
\end{figure}

The variance $\sigma(t)^2\equiv \la x^2(t)\ra-\la x(t)\ra^2$ is obtained as,
\begin{align}\sigma^2(t)=\dfrac{2A(1+\theta)^2}{(1+A)^4}\left[e^{-(1+A)t} +t+A t-1 \right].
\end{align}
Therefore, similar to the usual RTP, the variance shows two dynamical regimes 
depending on the value of $\tau_{\text{RT}}\equiv \min(1,A^{-1})$,  
\begin{align}
\sigma^2(t)\approx\begin{cases}
v_{\text{eff}}^2t^2 &\qquad \text{  for  }t\ll\tau_{\text{RT}},\\
2D_{\text{RT}}t&\qquad \text{  for  }t\gg \tau_{\text{RT}}
\end{cases}\ .
\label{genrtp:var_as}
\end{align}
where $ v_{\text{eff}}$ and $D_{\text{RT}}$ denote the effective velocity of the fluctuations,
\begin{align}
v_{\text{eff}}=\sqrt{2A}\frac{(1+\theta)}{(1+A)},
\end{align}
and the diffusion coefficient 
\begin{align}
    D_{\text{RT}}= A\frac{(1+\theta)^2}{(1+A)^3},
\end{align}
respectively.
Figure~\ref{fig:rtp:msd}a discusses $\sigma^2(t)$ for different values of $A$ showing the ballistic and diffusive regimes of the variance. Figure~\ref{fig:rtp:msd}b displays $ v_{\text{eff}}$ and $D_{\text{RT}}$
as functions of $A$. Both are zero in the limits $A\to 0$ and $A\to\infty$, as in these limits, the particle is almost deterministic with very few tumbling events. However, at the intermediate values of $A$, $v_{\text{eff}}$ and $D_{\text{eff}}$ reach a maximum at $A=1$ and $A=1/2$, respectively. The magnitude of the relative propulsion speed $\theta$ does not 
influence the position of the maximum; %
however, it determines their respective maximum values. 

\begin{figure}
    \centering
    \includegraphics[width=\hsize]{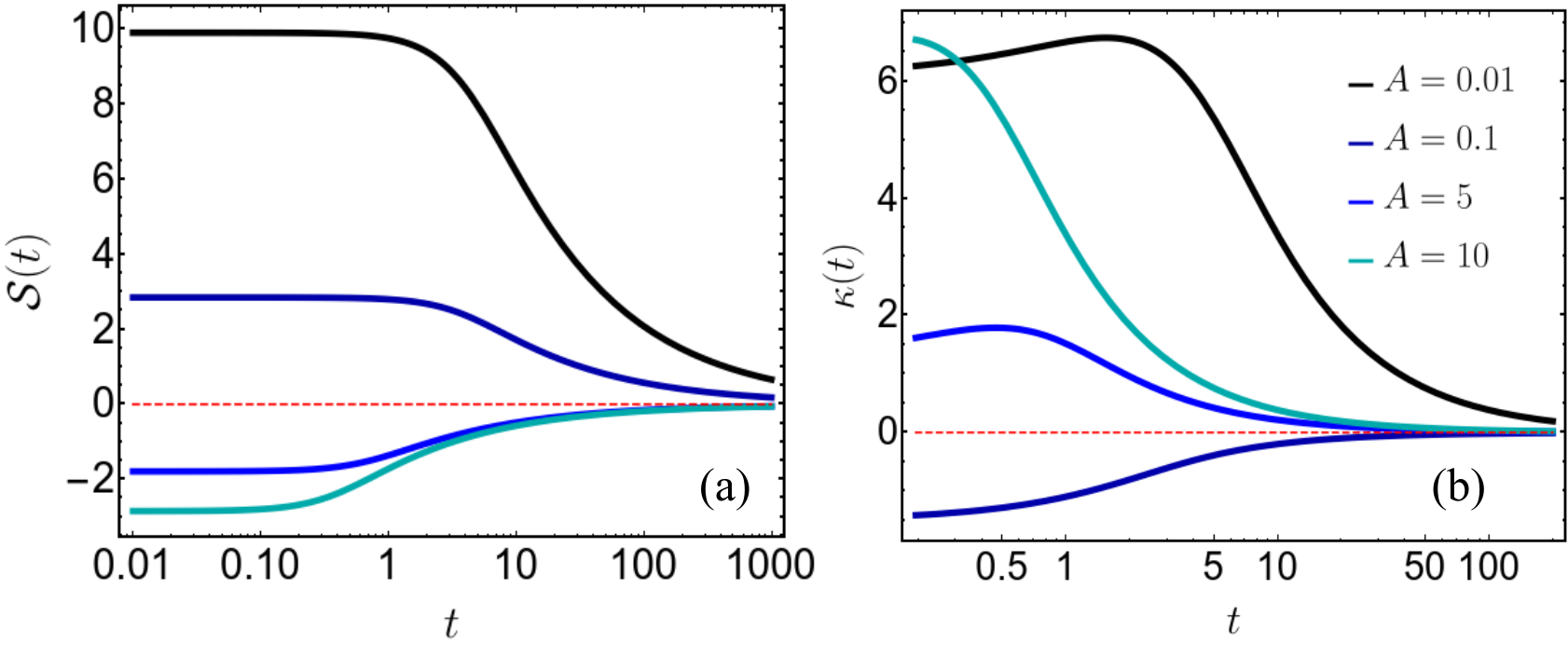}
    \caption{Approach of the position fluctuations of generalized run and tumble particle to a Gaussian distribution. Skewness (a) and kurtosis (b) as a function of time $t$. Color intensity increases with increasing the value of $A$. We fix the speed asymmetry in backward to forward motion to $\theta=1$. }
    \label{fig:rtp:sk-kurt}
\end{figure}

We further compute the skewness, $\mathcal{S}(t)=\frac{\la (x-\la x\ra)^3\ra}{\sigma(t)^{3}}$, of the position distribution, which quantifies the asymmetry of a distribution with respect to its mean value. The skewness vanishes at long-time, with the leading order behaviour,
\begin{align}
    \mathcal{S}(t)\approx \frac{3(1-A)}{\sqrt{2(A+A^2)}}\,t^{-1/2}.\label{gen:rtp:skew}
\end{align}
 The distribution becomes symmetric about its mean from a positively or negatively skewed distribution depending on $A>1$ or $A<1$. This is because for $A<1$, the time spent in the positive velocity phase is lesser compared to the time spent in the negative velocity phase. \textcolor{black}{This is shown in Fig.~\ref{fig:rtp:sk-kurt}a for different values of $A$.}
 
 Moreover, the excess kurtosis, $\mathcal{K}(t)=\frac{\la (x-\la x\ra)^4\ra}{\sigma(t)^{4}}-3$, can also be computed exactly and goes to zero as,
\begin{align}
    \mathcal{K}(t)\approx-\frac{6(1-4A+A^2)}{A(1+A)t}.\label{grtp:kurt:m}
\end{align}
The distribution thus tends to a Gaussian distribution from a platykurtic (negative kurtosis) or a leptokurtic side (positive kurtosis) for $2-\sqrt{3}<A<2+\sqrt{3}$, and $A\in (0,2-\sqrt{3})\cup(2+\sqrt{3},\infty)$, respectively, \textcolor{black}{as shown in Fig.~\ref{fig:rtp:sk-kurt}b}. For very small or large $A$, the particle spends more time moving in one direction, thus the distribution approaches the 
corresponding normal distribution from one which is more sharply peaked, leading to the positive kurtosis. The negative kurtosis for intermediate values of $A$, on the other hand, occurs due to the competition of the time spent in the forward and backward motion, which are of the same order, causing more fluctuations resulting in the approach to the corresponding Gaussian distribution to happen from a flatter side.

\subsection{Run-and-tumble particles with intermediate passive phases}\label{sec:rtpintermittent}
In the previous section~\ref{genrtp}, we discuss the case of an active particle in one dimension switching its directions (forward $\leftrightarrow$ backward), i.e., the generalized RTP. Herein, following the framework developed in Sec.~\ref{sec:general}, we extend our analysis to a case of a particle switching its motion between the generalized RTP and a passive (Brownian) dynamics.     
For convenience, we consider a case in which the particle is initialized in the generalized RTP phase (we refer to this phase as `phase 1'). The Fourier-Laplace transform of the propagator reads 
\begin{align}
   \tilde{\bar P}_1(k,s)=\tilde{\bar p}(k,s)\ , \label{eq1-pks}
\end{align}
where $\tilde{\bar p}(k,s)$ 
is given in \eqref{pks:active}.
`Phase 2', on the other hand, is 
a passive diffusion phase
with its propagator in the Fourier-Laplace space:
\begin{align}
    \tilde{\bar {P}}_2(k,s)=\frac{1}{s+Dk^2}\ .\label{eq2-pks}
\end{align}

The general formula for the Fourier-Laplace transform of the position distribution~\eqref{eq:fullgeneral} for waiting times being governed by Poisson distribution ($W_{ij}(t)\equiv r_{ij}e^{-r_{ij}t}$), is 
\begin{align}
     \tilde{\bar {P}}(k,s)=\frac{ \tilde{\bar {P}}_1(k,s+r_{\text{ap}})[1+r_{\text{ap}}  \tilde{\bar {P}}_2(k,s+r_{\text{pa}})]}{1-r_{\text{ap}} r_{\text{pa}}  \tilde{\bar {P}}_1(k,s+r_{\text{ap}}) \tilde{\bar {P}}_2(k,s+r_{\text{pa}})}\ .\label{rtp:diff:pks}
\end{align}
Thus, using the aforementioned $\tilde{\bar P}_1(k,s)$~\eqref{eq1-pks} and $\tilde{\bar P}_2(k,s)$~\eqref{eq2-pks} we 
find 
$\tilde{\bar {P}}(k,s)$ exactly~[see Appendix~\ref{s:rtp-pass}]. 
The expression of $\tilde{\bar {P}}(k,s)$ is rather long, and we relegate it to the Appendix~\ref{s:rtp-pass}. Further, it is also difficult to invert this Fourier-Laplace transformed expression. Nonetheless, we compute the first four cumulants to understand the statistics of this switching process (i.e., generalized RTP $\leftrightarrow$ Brownian dynamics). 

Employing~\eqref{momentsgen} for Eq.~\eqref{rtp:diff:pks}, we compute the first position moment:
\begin{align}
    \la x(t)\ra=\frac{(A-\theta)}{(1+A)}\dfrac{r_{\text{ap}}[1-e^{-(r_{\text{ap}}+r_{\text{pa}})t}]+r_{\text{pa}}(r_{\text{ap}}+r_{\text{pa}})t}{(r_{\text{ap}}+r_{\text{pa}})^2}\ .\label{mean:rtp-pas}
\end{align}
The mean position is independent of the diffusion coefficient, $D$, of the passive phase, since the only drift comes due to 
the particle's active motion. 

At short-time, the mean position retains exactly the same property
as the generalized RTP \eqref{g:rtp:mean}:
\begin{align}
   \lim_{t\to0}  \la x(t)\ra=\frac{(A-\theta)}{(1+A)}t.\label{mean:short-t}
\end{align} 
This can be understood as we consider that the particle always begins in the active phase, and in the short-time regime, $t\ll (r_{\text{ap}}$, there are typically no switching events.
In the long-time regime, though there have been appreciable number of switching events, the mean is the same as Eq.~\eqref{mean:short-t}, however, now weighed by the average time spent in the active phase:
\begin{align}
   \lim_{t\to\infty}  \la x(t)\ra=\frac{r_{\text{pa}}}{r_{\text{ap}}+r_{\text{pa}}}\frac{(A-\theta)}{(1+A)}t\ .
\end{align}

The variance or the second cumulant captures the position fluctuations.
In the following, we discuss its  asymptotic behaviours [however, the full-time dependent form can be calculated using \eqref{momentsgen} for Eq.~\eqref{rtp:diff:pks}]:
\begin{align}
    \sigma^2(t)\approx\begin{cases}
        \left[D r_{\text{ap}}+\frac{A  (v+1)^2}{(A +1)^2}\right]t^2  \quad & t\ll \min\{1,A^{-1}\,r_{\text{ap}^{-1}}\}\\
        \\
        2D_{\text{eff}}t \quad & t\gg \max\{1,A^{-1},r_{\text{ap}^{-1}}\},
    \end{cases}
    \label{var:ex}
\end{align}
where the effective diffusive coefficient is 
\begin{align}
\label{deff-eqn}
    D_{\text{eff}}&=\frac{D r_{\text{ap}}}{r_{\text{ap}}+r_{\text{pa}}}+\frac{r_{\text{pa}} (A +1) r_{\text{ap}}^2 \left(A +\theta^2\right)+A  r_{\text{pa}}^2 (\theta+1)^2}{(A +1)^2 (A +r_{\text{ap}}+1) (r_{\text{ap}}+r_{\text{pa}})^3}\nonumber\\
  &+\frac{r_{\text{pa}} r_{\text{ap}} \left[A^3+2 A  r_{\text{pa}} (\theta+1)^2+\theta^2+A ^2 (1-2 \theta)+A  (\theta-2) \theta\right]}{(A +1)^2 (A +r_{\text{ap}}+1) (r_{\text{ap}}+r_{\text{pa}})^3}\ .
\end{align}

\begin{figure}
    \centering\includegraphics[width=\hsize]{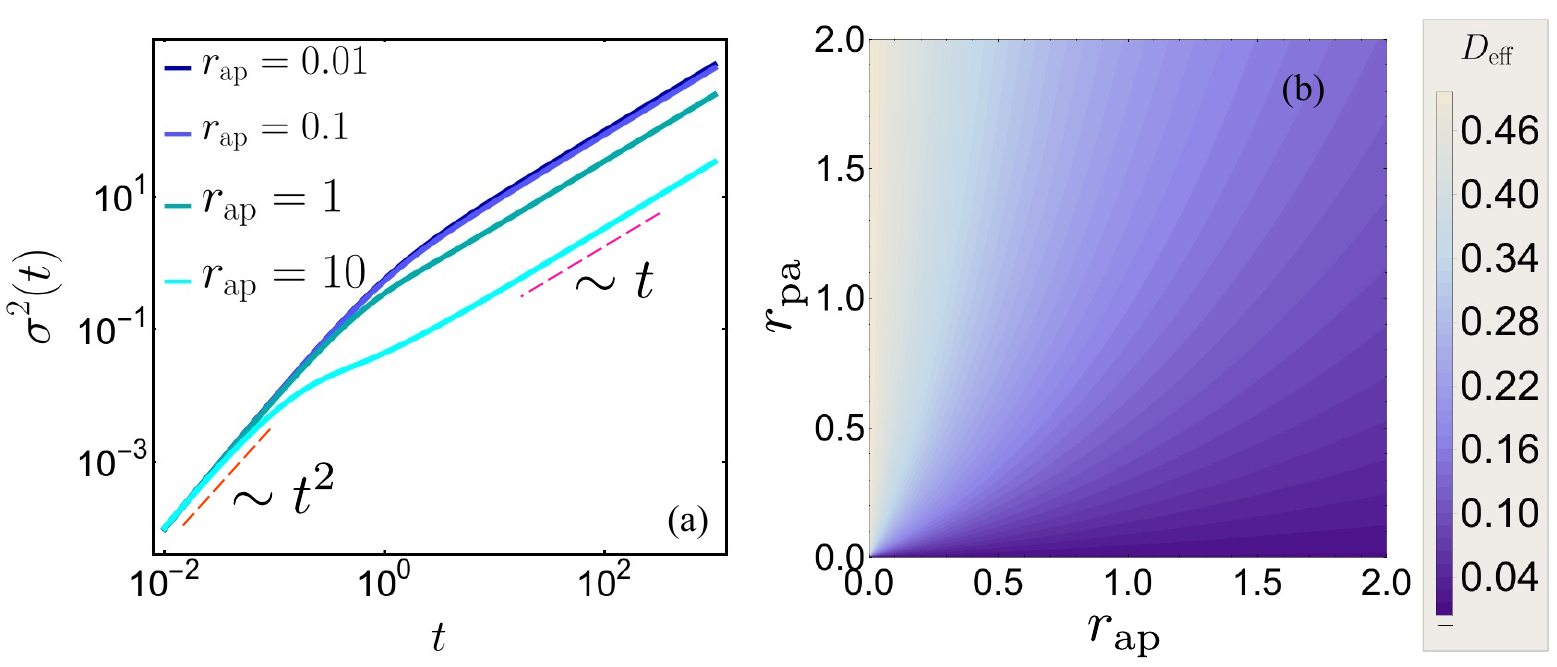}
    \caption{Position fluctuation of a generalize run-and-tumble particle with intermediate passive phases. (a) Variance as a function of time $t$. The color intensity increases with active to passive switching rate $r_{\rm ap}$. We fix  $r_{\text{pa}}=1$. 
    The red and magenta dashed lines, respectively, indicate the short-time ballistic and  long-time diffusive behaviour. (b) Effective diffusion coefficient 
    in the $(r_{\text{ap}},r_{\text{pa}})$ plane. Here, we fix $A=1$ and $D=0.01$.}
    \label{fig:rtp-pass:msd}
\end{figure}

Figure~\ref{fig:rtp-pass:msd}a displays the  time evolution of the variance for different values of $r_{\text{ap}}$, indicating the short-time ballistic and long-time diffusive regime [see Eq.~\eqref{var:ex}]. 

To 
investigate how the effective diffusion coefficient $D_{\rm eff}$~\eqref{deff-eqn} changes with respect to the switching rates, we 
analyse $D_{\text{eff}}$ in the $(r_{\text{ap}},r_{\text{pa}})$ plane. To this end, we note that, in the limit  $r_{\text{ap}}\to 0$, the effective diffusion coefficient $D_{\text{eff}}\to D_\text{RT}$. Moreover, since we consider the diffusion coefficient in the active phase $D_\text{RT}$ is larger than the diffusive phase (so that the diffusive phase is slower than the active phase), the effective diffusion coefficient of the particle is always peaked at the line $r_{\text{ap}}=0$. For finite values of $r_{\text{ap}}$, $D_{\text{eff}}$ increases with the increase in $r_{\text{pa}}$ as the time spent in the diffusive phases decreases. This is illustrated in Fig.~\ref{fig:rtp-pass:msd}b.

\begin{figure}
    \centering
    \includegraphics[width=\hsize]{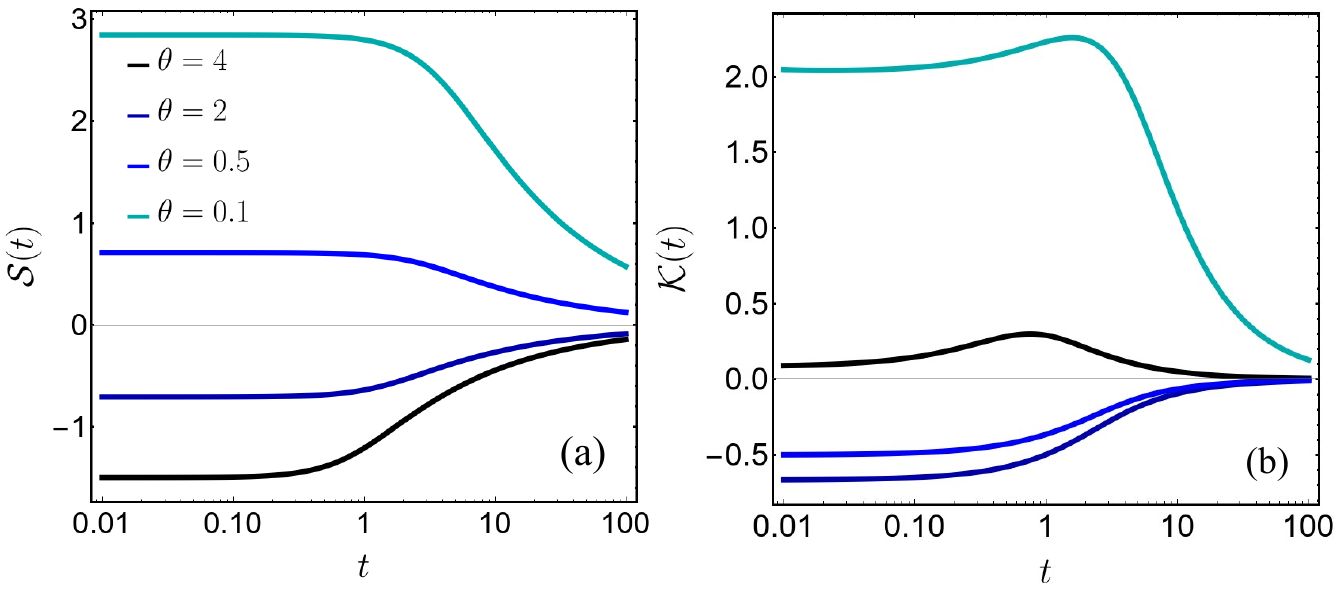}
    \caption{
    Approach of the position fluctuations of generalized run and tumble particle to a Gaussian distribution with intermediate passive phases.
    Skewness (a)  and kurtosis (b)  as a function of time $t$. 
    Here, we fix $r_{\text{pa}}=0.01$, $r_{\text{ap}}=1$, $D=0.01$ and different values of $\theta=A$. }
    \label{fig:rtp-pass:sk-ku}
\end{figure}

For the higher cumulants, 
we specifically focus on the case where $A=\theta$
(unbiased limit of the generalized RTP).  
It turns out that the skewness goes to zero 
in the long-time limit as
 \begin{align}
    \mathcal{S}(t)=\frac{3 r_{\text{pa}} (\theta-1) \theta \sqrt{r_{\text{ap}}+r_{\text{pa}}}}{\sqrt{2} \sqrt{r_{\text{ap}}+\theta+1} (D r_{\text{ap}} (r_{\text{ap}}+\theta+1)+r_{\text{pa}} \theta)^{3/2}} t^{-1/2},
\end{align}
indicating 
the resulting distribution to be a symmetric one. 
Depending on the case, if $\theta$ is smaller or larger than unity, the distribution becomes symmetric from a positively
or a negatively skewed distribution, respectively (Fig.~\ref{fig:rtp-pass:sk-ku}a). 
Thus, though  
temporal-scaling 
remain unchanged compared to the generalized RTP model [c.f.~\eqref{gen:rtp:skew}], the prefactor is modified.

The kurtosis can also be exactly computed and the leading order behaviour at long-time limit is given by,
\begin{align}
    \mathcal{K}(t)&\approx  t^{-1}\times\frac{6}{(r_{\text{ap}}+r_{\text{pa}}) (r_{\text{ap}}+\theta+1) (D r_{\text{ap}} (r_{\text{ap}}+\theta+1)+r_{\text{pa}} \theta)^2} \cr 
    &\Big[-D^2 r_{\text{ap}} (r_{\text{ap}}-r_{\text{pa}}) (r_{\text{ap}}+\theta+1)^3+D r_{\text{ap}} \theta (r_{\text{ap}}+\theta+1)\cr
  &  \left(r_{\text{ap}}^2+r_{\text{ap}} (-5 r_{\text{pa}}+\theta+1)-r_{\text{pa}} (2 r_{\text{pa}}+3 \theta+3)\right)\cr
  &+r_{\text{pa}} \theta \big(r_{\text{ap}}^2 \left(\theta^2+1\right)+2 r_{\text{ap}} (r_{\text{pa}} (\theta-3) \theta+r_{\text{pa}}+\theta^2+\theta)\cr
  &+r_{\text{pa}}^2 ((\theta-4) \theta+1)\big)\Big]\ .
    \end{align}
The $t^{-1}$--coefficient's sign
determines if the kurtosis approaches Gaussian from a platykurtic or leptokurtic distribution. For very small or large $\theta$, the particle when in the active phase spends more time in moving in one of the run phases, leading to sharper distribution than the corresponding Gaussian distribution at long-time; leading to $\mathcal{K}(t)$ approach zero from the positive side. For intermediate $\theta$, on the other hand, the competition between the forward and backward runs are stronger, leading to stronger position fluctuations causing the distribution to approach Gaussianity from the negative side. This is shown in Fig.~\ref{fig:rtp-pass:sk-ku}b for different values of $\theta$.

\section{Time evolution of the position distribution}
\label{sec:dist}
In the previous section, we discussed the cumulants of the two different switching process. Here, we analyse
the time-dependence of the density of a large number of non-interacting active particles 
with intermittent passive phase. 
For simplicity, we 
specialized to the case of the standard RTP (i.e., $A=\theta=1$) with $r_{\text{ap}}=r_{\text{pa}}=r$. The Fourier-Laplace transform of the probability distribution \eqref{rtp:diff:pks} in this case has a simpler form:
\begin{align}
    \tilde{\bar {P}}(k,s)=\frac{(r+s+2) \left(D k^2+2 r+s\right)}{D k^2 \left[k^2+(r+s) (r+s+2)\right]+k^2 (r+s)+s (r+s+2) (2 r+s)}.\label{pks:dist}
\end{align}
Since the above~\eqref{pks:dist} Fourier-Laplace transform is difficult to invert, we first discuss the time evolution behavior of the position distribution as seen from numerical simulations. Thereafter, we perform some asymptotic analysis.

Let us first discuss the time evolution of the position distribution qualitatively. From Fig.~\ref{dist:evolution}, we see that at short-time $t\ll \{1,r^{-1}\}$, the distribution has two different dynamical regions characterized by an RTP like behavior till $x=\pm t$, followed by non-trivial tail behavior, which we find asymptotically in Eq.~\eqref{tail}; at $x=\pm t$, there are two Dirac-delta function-like peaks which can be attributed to the particles that have not changed propulsion states, or switched to passive state, in time $t$. With the increase in time, this peak vanishes, and the two dynamical regions combine, eventually becoming Gaussian-like at long-time, as also predicted from the cumulant analysis in the previous sections, see Fig.~\ref{fig:rtp:sk-kurt}. In the following, we also find the asymptotic distribution at very long-time, starting from Eq.~\eqref{pks:dist}.

\subsection{Short-time distribution}

{\color{black}

To understand the short-time distribution $t\ll(r^{-1},\alpha^{-1})$, we perform a trajectory-based calculation based on the number of tumbles and switches that the particle has undergone. We write the distribution as a perturbative series in the active-passive switching rate $r$ and the RTP's tumbling rate  $\alpha_\text{fb}=\alpha_\text{bf}=\alpha=1$:
\begin{align}
P(x,t)=\sum_{m,n=0}^{\infty}P^{(n,m)}(x,t)\ ,\label{dist:series}
\end{align}
where $P^{(n,m)}(x,t)$ denotes the contributions of trajectories with $n$ switches and $m$ tumbles and is of the order $\mathcal{O}(r^{n}\alpha^{m})$. For time $t\ll (\alpha^{-1},r^{-1})$, the first few terms in the above series is expected to provide a good description to the short-time distribution. Let us systematically compute the first few order contributions.

For trajectories that have undergone neither tumbling nor switching events, we have the following contribution:
\begin{align}
    P^{(0,0)}=e^{-(\alpha+r) t}\frac{1}{2}\sum_{j=\pm 1}\delta(x-jt)\ ,\label{eq:delta}
\end{align}
where we assumed that the particle starts from $x=0$ in the active phase $j=\pm 1$ with equal probability $1/2$. 
Notice that the above contribution~\eqref{eq:delta} is non-zero only at $x=\pm t$, and its weight decreases as time increases. Figure~\ref{dist:evolution}a and b demonstrates these delta functions at $x/t=\pm 1$.

Next for trajectories with one switch but no tumbling events, i.e., the contribution of $\mathcal{O}(r)$, we have
\begin{align}
    P^{(1,0)}(x,t)=\sum_{j=\pm 1}\frac 12\int_{-\infty}^{+\infty} dx_1\int_0^t dt_1\left[ re^{-(r+\alpha)t_1}\delta(x_1-jt_1)\right]\nonumber\\
    \times \left[\frac{e^{-r(t-t_1)}e^{-\frac{(x-x_1)^2}{4D(t-t_1)}}}{\sqrt{4\pi D(t-t_1)}}\right]\ .\label{eq:zero}
\end{align}
 The terms in the first squared brackets denote the probability that the particle, starting from $x=0$ in the active state $j$, does not undergo any tumbling or switching event till time $t_1$ and reaches to $x_1= jt_1$, then, it switches to the passive-phase. The terms in the second squared brackets denotes the probability that once the particle in the passive-phase it stays in the same phase in the remaining time $t-t_1$ and reaches to the position $x$. The integration~\eqref{eq:zero} over $x_1$ gives
\begin{align}
    P^{(1,0)}(x,t)=\sum_{j=\pm 1}\frac 12 r e^{-rt}\int_0^t~dt_1 e^{-\alpha t_1}\frac{e^{-\frac{(x-jt_1)^2}{4D(t-t_1)}}}{\sqrt{4\pi D(t-t_1)}}\quad\text{for }|x|>t\ .
\end{align}
An exact closed form of the above integral cannot be obtained. Nevertheless, the integration can be performed numerically. 

Next we consider terms of order $\mathcal{O}(r \alpha )$. This comes from trajectories which have undergone one tumbling event at time $t_1$ before switching to the passive-phase at time $t_1+t_2$, 
\begin{align}
    P^{(1,1)}(x,t)&=\sum_{j=\pm 1}\frac 12\int_{-\infty}^{+\infty} dx_1\int_0^t dt_1\int_0^{t-t_1} dt_2\big[\alpha r e^{-(r+\alpha)(t_1+t_2)}\nonumber\\
    &\times\delta(x_1-j(t_1-t_2))\big]
     \left[\frac{e^{-r(t-t_1-t_2)}e^{-\frac{(x-x_1)^2}{4D(t-t_1-t_2)}}}{\sqrt{4\pi D(t-t_1-t_2)}}\right]\ ,\label{eq:one}
\end{align}
where $x_1$ denotes the distance travelled till time $t_1+t_2$, and the second brackets correspond to the particle staying in the passive phase in the remaining time $t-t_1-t_2$.
The integral~\eqref{eq:one} over $x_1$ can be performed, and it gives
\begin{align}
    P^{(1,1)}(x,t)&= \frac {\alpha r e^{-rt}}{2}\sum_{j=\pm 1}\int_0^t dt_1\int_0^{t-t_1} dt_2\left[\frac{e^{-\alpha(t_1+t_2)}e^{-\frac{(x-j(t_1-t_2))^2}{4D(t-t_1-t_2)}}}{\sqrt{4\pi D(t-t_1-t_2)}}\right]\ .
\end{align}

Now let us evaluate the contribution from trajectories, which have undergone tumbling events, but have never switched to the passive phase. Such contributions $P^{(0,1)}(x,t)$ is given by
\begin{align}
P^{(0,1)}(x,t)=\frac{e^{-r t}}{2}\sum_{j=\pm 1}\int_0^t dt_1\alpha e^{-\alpha t_1}e^{-\alpha (t-t_1)}\delta\left(x-j [t_1-(t-t_1)]\right).
\end{align}
The right-hand side of above equation can be understood as follows: $e^{-r t}$ accounts for no switching in the entire duration $t$, the factor $\alpha e^{-\alpha t_1}$ denotes that the particle undergoes the first tumble event at $t_1$; $e^{-\alpha (t-t_1)}$ denotes that the particle does not undergo any further tumbling event in the remaining time $t-t_1$, and finally the delta function denotes the position of the particle having started in a state $j$. Performing the integral over $t_1$, we get
\begin{align}
    P^{(0,1)}(x,t)&=\frac{\alpha e^{-(\alpha+r) t}}{2}\Theta(t-|x|)\ .\label{inn-one}
\end{align}
Similarly, for two tumbles (and no switching event), the contribution to the probability distribution is
\begin{align}
P^{(0,2)}(x,t)&=\frac{\alpha^2 e^{-(\alpha+r) t}}{4}\Theta(t-|x|)\ .\label{inn-two}
\end{align}

Analogously, one can find higher-order terms systematically. However, to understand the observed distributions we truncate the series~\eqref{dist:series} for short-time as
\begin{align}
    P(x,t)=P^{(0,0)}(x,t)&+P^{(0,1)}(x,t)+P^{(0,2)}(x,t)+P^{(1,0)}(x,t)\cr
    &+P^{(1,1)}(x,t)+\mathcal{O}(\text{min}[r^2,\alpha^3])\ .\label{eq:approx}
\end{align}
Note that the Heaviside theta functions in both $P^{(0,1)}$ and $P^{(0,2)}$ indicate that the contributions from these trajectories remain confined with $|x|< t$. Thus, the contribution to the distribution $|x|>t$ comes from trajectories with at least one switch, and for short-time the tails' distribution can be approximated as 
\begin{align}
    P_{\text{tail}}(x,t)\approx P^{(0,1)}(x,t)+P^{(0,2)}(x,t)\label{tail}\ .
\end{align}
Figure~\ref{dist:evolution} shows the comparison of tails' analytical results~\eqref{tail} with the numerical simulations, and it shows a good agreement.

The inner region $|x|<t$, however, requires the contribution from all the terms on the right-hand side of Eq.~\eqref{eq:approx}. 
Figures~\ref{dist:evolution}a and b show the comparison of analytical results [Eqs.~\eqref{eq:approx} and~\eqref{tail}] with the numerical simulation for inner-region. We find that $\alpha^{-1}<t<r^{-1}$, we see that the inner region shows a deviation. This is due to the fact that the particle's trajectories have undergone multiple tumbles without any switching events, and higher-order terms in Eq.~\eqref{eq:approx} are required. 
}

\subsection{Long-time distribution}
\begin{figure}
    \centering
    \includegraphics[width=\hsize]{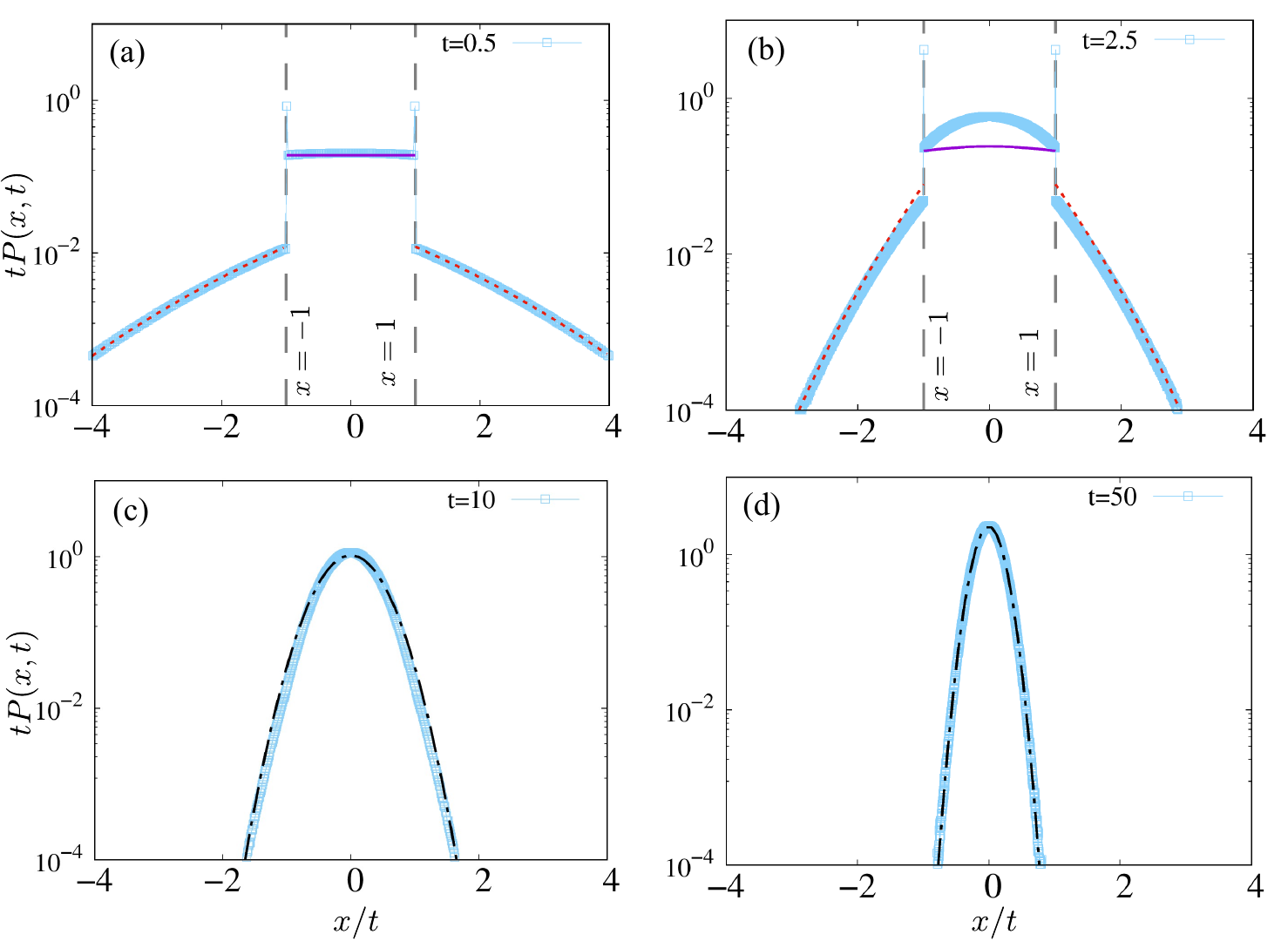}
    \caption{\textcolor{black}{Time evolution of the position distribution of the standard run-and-tumble particle with intermittent passive phases. Symbols: Numerical simulations. (a) and (b) 
    correspond to short-time distributions for $t\ll (r^{-1},\gamma^{-1})$, and $\gamma^{-1}<t<r^{-1}$ respectively. The red dashed lines denote the short-time analytical prediction~\eqref{tail} for the tails, while the solid purple lines denote Eq.~\eqref{eq:approx}. The dashed vertical lines denote the position of the delta function~\eqref{eq:delta}. The dot-dashed lines in (c) and (d) denote the large-time Gaussian behavior~\eqref{dist:lt}.
    Here, we fix $D=1$ and $r=0.1$.}}
    \label{dist:evolution}
\end{figure}

The behaviour of the distribution near the tails are dominated by contributions coming from small-$k$ behaviour. In this limit, the Fourier-Laplace transform simplifies Eq.~\eqref{pks:dist} to,
\begin{align}
\label{pks-approx1}
    \tilde{\bar {P}}(k,s)\approx \frac{(r+s+2) \left(2 r+s\right)}{(r+s+2) \left[(r+s) \left(D k^2+s\right)+s\right]+k^2 (s+1)}\ .
\end{align}
Additionally, in the limit $s\to 0$, which corresponds to  the long-time behaviour, the above equation~\eqref{pks-approx1} simplifies to:
\begin{align}
    \tilde{\bar {P}}(k,s)\approx \frac{1}{s+k^2[D+(2+r)^{-1}]/2}\ ,
\end{align}
which by inverting the Fourier and Laplace transform, leads to a diffusive Gaussian scaling 
of the tails at long-time:
\begin{align}
    P(x,t)=\dfrac{1}{\sqrt{2\pi t [D+(2+r)^{-1}]}}\exp\left(-\frac{x^2}{2t[D+(2+r)^{-1}]}\right)\ .\label{dist:lt}
\end{align}
 \textcolor{black}{
 Figures~\ref{dist:evolution}c and d show a good agreement of the Gaussian description~\eqref{dist:lt} with the numerical simulation for longer times ($t\gg r^{-1}$).}

\section{Summary}
\label{sec:summ}
We analyzed a dynamical process characterized by a stochastic switching mechanism. By employing a trajectory-based approach, we derived the Fourier-Laplace transform of the position distribution for a particle undergoing such switching dynamics, with arbitrary waiting times. We calculated the generalized diffusion coefficient for this system, assuming the waiting time distributions possess a finite mean. We then applied this framework to an example of an active particle exhibiting intermittent diffusive phases, where we computed the exact expressions for the first few position cumulants. Additionally, we derived asymptotic expressions for the tails of the probability density function at all times, and our results were corroborated by numerical data.     

Although our study focuses on one-dimensional processes, the general formulas we developed are applicable to any number of dimensions and can be utilized to investigate switching processes in higher-dimensional settings. The first-passage properties of these active processes display intriguing characteristics, particularly when compared to standard diffusion~\cite{basu2023target}. It would be interesting to explore whether intermittent diffusive phases could potentially enhance the first-passage times of active particles. {\color{black}Furthermore, another promising direction for future research is to examine the role of 
confining potential~\cite{pototsky2012active}, many interacting particles, and 
stochastic resetting~\cite{evans2020stochastic} on the 
distribution of switching processes.}
{\color{black} Finally, there are multiple intriguing extensions to more than two dynamical phases. Processes that switch between $N$ dynamical phases can often be decomposed into multiple two-phase systems. For such decomposition, one has to modify the rates of switching from one phase to another and vice-versa in the effective sense~\cite{Yuhai-Tu}. This is an interesting topic, and could be considered for future research. In the future, repeated application of Eq. (\ref{eq:fullgeneral}) could be used to study such higher-order switching dynamics. In such cases, one could also consider cases where switching is coupled to the system's state, e.g., by only allowing switching transitions in a subset of the phases. Further, in this paper, we discussed switching mechanism by an instantaneous protocol, our results, however, can be generalized to non-instantaneous switching protocols by introducing an additional intermediate phase; such generalization and application of different temporal distribution of intermediate phase will be considered for future research.}


\section*{Acknowledgements}
K.S.O acknowledges support by the Deutsche Forschungsgemeinschaft (DFG) within the project LO 418/29-1. D.G acknowledges support from the  Nordita fellowship program. Nordita is partially supported by Nordforsk. {\color{black}The authors thank the anonymous reviewers for their valuable suggestions that have improved the content and presentation of the manuscript.}\\


\appendix



\section{Exact expressions used in the main text}\label{s:rtp-pass}

In this section we provide the exact expressions for some of the expressions whose asymptotes or derivatives have been shown in the main text, namely skewness and kurtosis for the generalized RTP supplementing Sec.~\ref{genrtp} and the generating function of the RTP with intermediate passive phases used in Sec.~\ref{sec:rtpintermittent}.

\subsection{Generalized RTP}

    Using the generating function Eq.~\eqref{pks:active} and \eqref{momentsgen}, the skewness of the position of the  generalized RTP is given as,
\begin{multline}
\mathcal{S}(t)=-\frac{3 (A-1) e^{-(A+1) t} \left(A t+e^{(A+1) t} (A t+t-2)+t+2\right)}{\sqrt{2} \sqrt{A} \left(A t+e^{-(A+1) t}+t-1\right)^{3/2}}
\label{grtp:skew:exact}
\end{multline}
which at large time, reduces to Eq.~\eqref{gen:rtp:skew}. 
Similarly, from the fourth moment of the position distribution, the kurtosis can be obtained as
\begin{align}
    \mathcal{K}(t)&=\frac{1}{A \left(e^{(A+1) t} (A t+t-1)+1\right)^2}\left(3 e^{2 (A+1) t}-3 A\right.\\
    &\left.\times(-6 A^2+2 (A+1) ((A-3) A+1) t+17 A-6)+3 e^{(A+1) t}\right.\cr
    &\left.\left(A^2-1\right)^2 t^2+4 (A+1) ((A-3) A+1) t+2 A (3 A-8)+6\right).\nonumber
    \label{grtp:kurt:ex}
\end{align}
The above equation at long-times reduces to Eq.~\eqref{grtp:kurt:m}

\subsection{RTP with intermittent passive phases}
The Fourier-Laplace transform of the distribution is given by,
\begin{align}
    \tilde{\bar {P}}(k,s)&=\left(D k^2+r_{\text{ap}}+r_{\text{pa}}+s\right) (k-A \theta k+i (A +1) (A +r_{\text{ap}}+s\cr
    &+1))/\Bigg[(\alpha +1) z \left(H \left(D k^2+r_{\text{pa}}+s\right)-r_{\text{ap}} r_{\text{pa}} (k-A  k \right.\cr
    &\left.+i (\alpha +1) (A +r_{\text{ap}}+s+1))\right)\Bigg].
\end{align}
Here,
\begin{align}
    H\equiv (A +1) (A  (k+i (r_\text{ap}+s))+(k+i (r_\text{ap}+s+1)) (i k \theta+r_\text{ap}+s)).
\end{align}

\balance



\providecommand*{\mcitethebibliography}{\thebibliography}
\csname @ifundefined\endcsname{endmcitethebibliography}
{\let\endmcitethebibliography\endthebibliography}{}

\end{document}